\documentclass[preprint,amsmath,amssymb,aps, superscriptaddress
]{revtex4-2}

\usepackage{booktabs}
\usepackage{siunitx}
\usepackage{graphicx}
\usepackage{stmaryrd}	
\usepackage{multirow}
\usepackage{soul}
\usepackage{makecell}
\usepackage{array} 
\usepackage{setspace}
\usepackage[table]{xcolor}

\newcommand{\BST}{Ba$_{1-x}$Sr$_{x}$TiO$_3$\ }

\begin{document}

\title{Compositionally tuned phase transformations enhance pyroelectric energy harvesting from low-grade heat}

\author{Ruiheng Geng}
\affiliation{Department of Mechanical and Aerospace Engineering, The Hong Kong University of Science and Technology, Hong Kong}

\author{Ka Hung Chan}
\affiliation{Department of Mechanical and Aerospace Engineering, The Hong Kong University of Science and Technology, Hong Kong}
\affiliation{Advanced Light Source, Lawrence Berkeley National Laboratory, Berkeley, CA, United States}

\author{Xinyue Huang}
\affiliation{Department of Mechanical and Aerospace Engineering, The Hong Kong University of Science and Technology, Hong Kong}

\author{Nobumichi Tamura}
\affiliation{Advanced Light Source, Lawrence Berkeley National Laboratory, Berkeley, CA, United States}

\author{Faqiang Zhang}
\affiliation{State Key Laboratory of High Performance Ceramics, Shanghai Institute of Ceramics, Chinese Academy of Sciences, 585 Heshuo Road, Shanghai 201899, China}

\author{Wanjia Han}
\affiliation{State Key Laboratory of High Performance Ceramics, Shanghai Institute of Ceramics, Chinese Academy of Sciences, 585 Heshuo Road, Shanghai 201899, China}

\author{Yang Zhang}
\affiliation{MOE Key Laboratory of Advanced Micro-Structured Materials, School of Physics Science and Engineering, Institute for Advanced Study, Tongji University, Shanghai 200092, China}

\author{Chenbo Zhang}
\email{cbzhang@tongji.edu.cn}
\affiliation{MOE Key Laboratory of Advanced Micro-Structured Materials, School of Physics Science and Engineering, Institute for Advanced Study, Tongji University, Shanghai 200092, China}

\author{Xian Chen}
\email{xianchen@ust.hk}
\affiliation{Department of Mechanical and Aerospace Engineering, The Hong Kong University of Science and Technology, Hong Kong}

\begin{abstract}
Phase-transforming pyroelectric materials have emerged as promising candidates for low-grade thermal energy harvesting. However, whether first-order transformations with large pyroelectric coefficient or second-order transformations with better reversibility are preferable remains unclear. Here we report compositionally tunable phase transformations in Ba$_{1-x}$Sr$_x$TiO$_3$ ($x \in [0, 0.3]$), revealing evolution from first-order to second-order character. We identify a transitional regime between Sr$_{0.15}$ and Sr$_{0.22}$ where transformation mechanism fundamentally changes. Within this regime, Sr$_{0.19}$ achieves optimal lattice compatibility, exhibiting electrical leakage suppressed by over two orders of magnitude while retaining substantial polarization response. Energy conversion demonstrations show the multilayer Sr$_{0.19}$ device delivers pyroelectric current of $\sim$1.6 $\mu$A at 64$~^\circ$C with an energy density of 1.6 mJ/cm$^3$ per cycle and 5.5\% conversion efficiency. Remarkably, this composition operates stably over 10,000 full energy conversion cycles without external bias field or recharging, demonstrating that transitional regime compositions provide the optimal balance between energy density and operational durability for practical low-grade heat harvesting.
\end{abstract}

\maketitle

\noindent\textbf{Introduction}

The escalating demand for sustainable energy solutions has intensified interest in harvesting low-grade waste heat, which often exists as temperature fluctuations below 150$~^\circ$C in industrial processes and residential environments. Conventional thermal energy harvesting approaches such as thermoelectric generators (TEG) and organic Rankine cycle (ORC) systems often struggle with much reduced power density at low-temperature regime, and deployment complexity \cite{ji2023comparison, he2024advances, quoilin2013techno}. Pyroelectric energy conversion (PEC), which exploits the temperature-dependent polarization of ferroelectric materials, offers an alternative approach for converting low-grade heat (i.e. 100 -- 200$~^\circ$C) to electricity directly \cite{bowen2014, zhang2019power, lee2013, bucsek2020energy, sebald2009, zhang2021energy, zhang2023longcycle}. Some works even demonstrated heat to electricity conversion below $100^\circ$C \cite{pandya2018, lheritier2022, zhang2025enhanced}, but the generated pyroelectric current was reported to be very small. Traditional pyroelectric energy conversion devices operate via the Olsen cycle \cite{olsen1985pyroelectric}, comprising two isothermal processes and two isobaric (constant voltage) processes, where heat is absorbed and released during the isobaric processes accompanied by electrical energy output \cite{lee2012pyroelectric, pandya2018, lheritier2022large}. However, this method requires significant DC bias fields to drive promising polarization changes, presenting a fundamental paradox: the thermodynamic cycle relies on external electrical input to harvest electrical energy from waste heat. As the temperature-dependent polarization changes in conventional ferroelectrics are too subtle, making it impractical to achieve meaningful energy conversion without external field.

To address this limitation, a power-source-free pyroelectric energy conversion using phase-transforming ferroelectrics was proposed \cite{zhang2019power, zhang2021energy, zhang2023longcycle}. By this approach, the ferroelectric capacitor is initially charged, and then the external bias source is disconnected during thermodynamic cycles, allowing the material's intrinsic phase transformation to drive large polarization jump without any attached DC voltage sources. Utilizing the first-order phase transformations is the key to boost electricity generation from thermal fluctuations. A new figure of merit (FOM) was proposed to underlie the performance of the transforming ferroelectric materials, which guides the material development for PEC devices \cite{zhang2019power, zhang2021energy, huang2024apl}. In addition, practical energy harvesting devices must sustain thousands of thermal cycles without performance degradation. The reversibility of phase transformation becomes critical. Symmetry-breaking structural transitions in perovskite ferroelectrics lead to microstructure incompatibility \cite{zhang2021low,zhang2023longcycle, huang2024apl} between paraelectric and ferroelectric phases. Moreover, electrical leakage presents another challenge for long-term operation, as elevated temperatures during thermal cycling can significantly increase leakage current density, leading to charge dissipation and reduced energy output \cite{zhang2020leakage}. This presents a multifaceted materials design challenge: achieving compositions that not only exhibit large polarization changes for high FOM, but also maintain lattice compatibility and low leakage current density. 

Compositional doping in perovskite ferroelectric materials is an effective method to manipulate phase transformation behavior. In the BaTiO$_3$ system, A-site substitution (barium replacement) directly influences transformation temperature and lattice parameters, thereby controlling structural compatibility. Ca$^{2+}$ doping is commonly used to manipulate transition temperature and latent heat. However, at higher doping compositions, it often induces diffuse or relaxor-like characteristics, which broaden the phase transformation region and reduce the pyroelectric coefficient, thus lowering the figure of merit (FOM) \cite{victor2003normal, khan2022role}. Trivalent rare earth dopants such as La$^{3+}$ and Sm$^{3+}$ maintain sharp transitions but increase free-carrier concentration, deteriorating insulation and aggravating leakage currents \cite{ganguly2013La, ganguly2013Sm}. Compared with other dopants, Sr$^{2+}$ is an isovalent dopant that reduces transformation temperature linearly predicted by FerroAI model \cite{zhang2025ferroai} while maintaining charge neutrality and low leakage \cite{Lemanov1996}. As a single-parameter compositional variable, Sr$^{2+}$ doping offers excellent tuning accuracy with direct correlation to lattice parameters. This makes Sr$^{2+}$ an ideal choice for investigating phase transformation properties in the BaTiO$_3$ system to achieve optimal energy conversion performance.

In this work, we systematically investigate the phase transformation behavior of barium titanate with Sr$^{2+}$ doping, which underlie the energy conversion performance from heat to electricity. As the Sr composition increases, we observe a transition in phase transformation behavior from first-order to second-order, accompanied by changes in thermal, structural symmetry, and ferroelectric properties. Through advanced structural analysis by synchrotron X-ray diffraction, we examined the compatibility conditions for the (Ba,Sr)TiO$_3$ material system and identified the optimal composition associated with the lowest thermal hysteresis and highest phase reversibility upon thermal cycles for energy conversion. 

\noindent\textbf{Results and Discussion}

\noindent\textbf{Microstructure and phase transformation behavior}

\begin{figure}[h]
\centering
\includegraphics[width=1\textwidth]{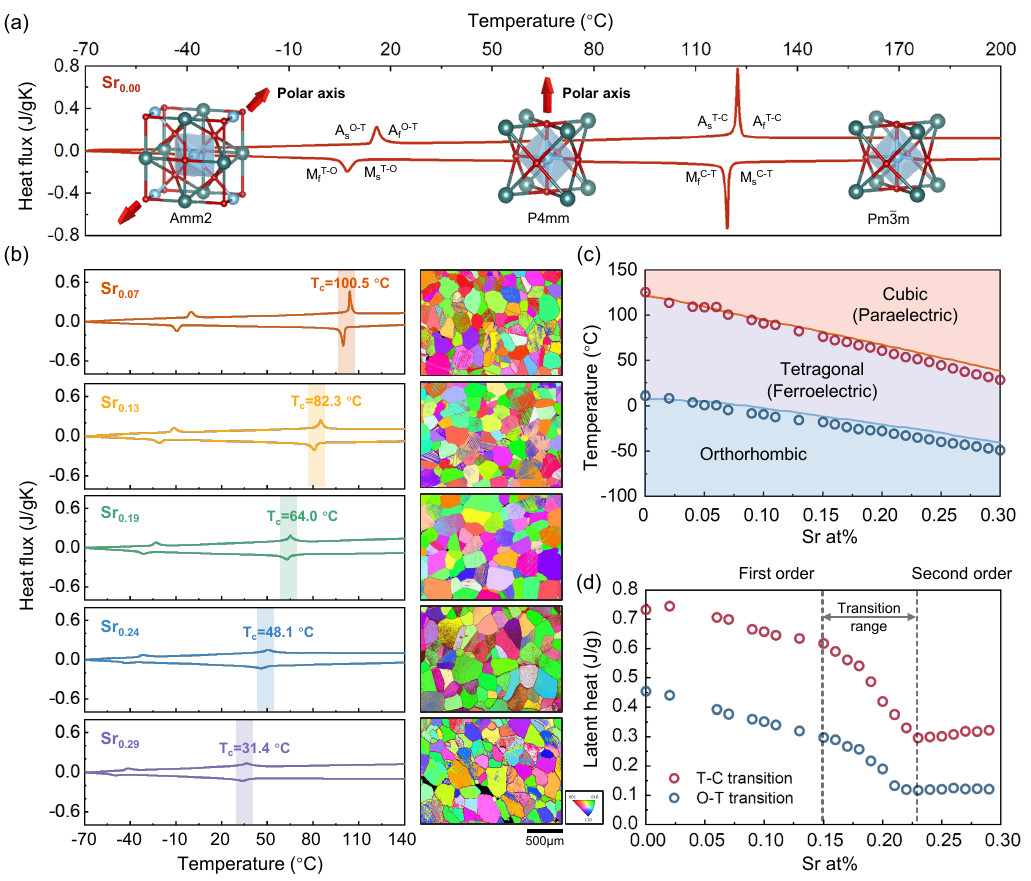}
\caption{\label{fig:latent heat} DSC heat flow curves for (a) Sr$_0$, (b) Sr$_{0.07}$, Sr$_{0.13}$, Sr$_{0.19}$, Sr$_{0.24}$ and Sr$_{0.29}$ samples associated with EBSD orientation maps of grain morphology. (c) Phase diagram of the Ba$_{1-x}$Sr$_x$TiO$_3$ system showing phase transformation temperatures as a function of Sr composition. Lines represent predictions from FerroAI and circles represent experimental measurements from DSC characterization, demonstrating good agreement between model and experiment. (d) Latent heat as a function of Sr composition for both cubic-to-tetragonal and tetragonal-to-orthorhombic phase transformations, revealing a transition in transformation order between Sr$_{0.15}$ and Sr$_{0.22}$.}
\end{figure}

The sintered specimens exhibit equiaxed grain morphology with randomly distributed orientations and an average grain size of approximately 100 $\mu$m, as revealed by electron backscatter diffraction (EBSD) (Figure \ref{fig:latent heat}). Sub-grain twinning structures were observed across the Sr compositions from 0 to 0.3 in \BST, suggesting phase transformation takes place from high-symmetry structure, i.e. cubic paraelectric phase. 

Differential scanning calorimetry reveals two reversible phase transformations within the temperature range of $-70~^\circ$C to $140~^\circ$C for all compositions: a cubic-to-tetragonal transition in the high-temperature regime and a tetragonal-to-orthorhombic transition in the low-temperature regime. Figure \ref{fig:latent heat}(a) and (b) show representative heat flow curves for compositions Sr$_{x}$ where $0\leq x \leq 0.29$.

We utilized FerroAI \cite{zhang2025ferroai}, a deep-learning model, to predict the phase diagram of the Ba$_{1-x}$Sr$_x$TiO$_3$ system with compositional variable $x \in [0, 0.3]$, as shown in Figure~\ref{fig:latent heat}(c). Within the temperature range from $-100^\circ$C to $150^\circ$C, three distinct phase regions are identified: the cubic paraelectric phase, the tetragonal ferroelectric phase, and the orthorhombic phase.

From the DSC measurements, the phase transformation temperature ($T_c$) can be determined as
\begin{equation}
T_c = \frac{M_s + M_f + A_s + A_f}{4},
\end{equation}
where $M_s$ and $M_f$ denote the start and finish temperatures during cooling, and $A_s$ and $A_f$ denote the start and finish temperatures during heating, respectively. The extracted $T_c$ values decrease linearly with increasing Sr content across all compositions, in remarkable agreement with the predicted phase diagram.

The latent heat, calculated from integrated heat flow peaks, exhibits distinctly different behavior as Sr composition varies, plotted in Figure \ref{fig:latent heat}(d). While transformation temperatures decrease linearly, latent heat shows an abrupt drop between Sr$_{0.15}$ and Sr$_{0.22}$ for both symmetry-breaking transformations. For Sr $<$ 0.15, latent heat is approximately 0.6 J/g (cubic-to-tetragonal) and 0.3 J/g (tetragonal-to-orthorhombic), characteristic of first-order transformations. Beyond Sr$_{0.22}$, latent heat reduces by a factor of two, indicating second-order character. The composition range between Sr$_{0.15}$ and Sr$_{0.22}$ represents a transitional regime where transformation mechanism evolves from first-order to second-order behavior.

This compositionally-induced shift in transformation order, occurring independently of equilibrium phase boundaries, reveals decoupling between thermodynamic phase stability and transformation kinetics. The identification of this evolution provides critical insight for optimizing pyroelectric energy conversion performance through targeted compositional design.

\noindent\textbf{Structural characterization by synchrotron X-ray diffraction}

\begin{figure}[h]
\centering
\includegraphics[width=0.85\textwidth]{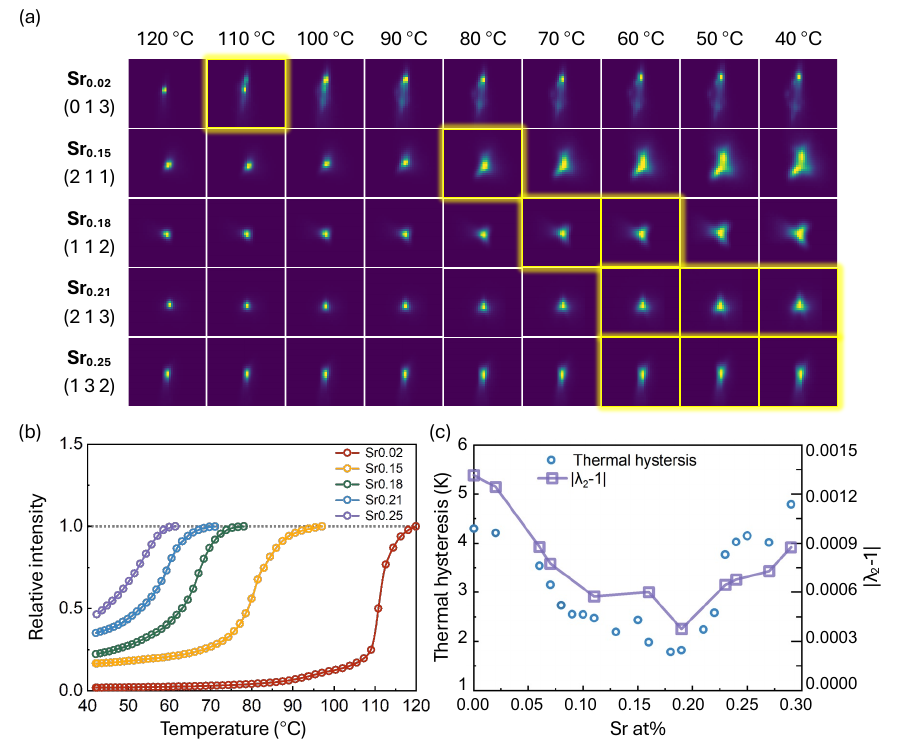}
\caption{\label{fig:laue diffraction}(a) Laue diffraction patterns of specific crystal planes at different temperatures in Sr$_{0.02}$, Sr$_{0.15}$, Sr$_{0.18}$, Sr$_{0.21}$ and Sr$_{0.25}$ samples. (b) Relative intensity of Laue pattern changes with temperature varying in  Sr$_{0.02}$, Sr$_{0.15}$, Sr$_{0.18}$, Sr$_{0.21}$ and Sr$_{0.25}$ samples. (c) Thermal hysteresis calculated from DSC results with $\lvert\lambda_2$-1$\rvert$ calculated from lattice parameter characterized by synchrotron XRD, ALS beamline 12.3.2.}
\end{figure}

Synchrotron X-ray diffraction measurements provide direct evidence for the evolution of the transformation mechanism. Laue diffraction patterns captured across the cubic-to-tetragonal transformation (Figure~\ref{fig:laue diffraction}(a)) reveal systematic changes in lattice distortion with composition. For the Sr$_{0.02}$ and Sr$_{0.15}$ samples, selected diffraction peaks exhibit pronounced splitting and elongation, indicating substantial lattice distortion characteristic of a first-order transformation. In contrast, the Sr$_{0.21}$ and Sr$_{0.25}$ samples show minimal spot splitting, reflecting reduced lattice distortion consistent with a second-order character. The peak shape evolution in Sr$_{0.18}$ is intermediate, suggesting a transitional transformation behavior.

Temperature-dependent intensity tracking of characteristic diffraction peaks quantifies this evolution (Figure \ref{fig:laue diffraction}(b)). Sr$_{0.02}$ exhibits sharp transitions within 5$~^\circ$C intervals, while the transition range progressively broadens from Sr$_{0.15}$ to Sr$_{0.25}$. This systematic broadening reflects the evolution from discontinuous first-order to continuous second-order character.

Lattice parameters of cubic, tetragonal and orthorhombic structures determined from monochromatic energy scans enable quantitative assessment of compatibility. The transformation stretch tensor between cubic and tetragonal phases is:
\begin{equation} \label{eq:U}
\mathbf{U}=\begin{bmatrix}
\frac{a_t}{a_c} & 0 & 0\\
0 & \frac{a_t}{a_c} & 0\\
0 & 0 & \frac{c_t}{a_c}
\end{bmatrix},
\end{equation}
where $a_t$ and $c_t$ are tetragonal lattice parameters and $a_c$ is the cubic parameter. The temperature-dependent lattice parameters for various symmetries are provided in Tables S1 and S2 of the Supplemental Materials. Figure S1 shows the trend of tetragonality as a function of temperature across the ferroelectric-to-paraelectric phase transformation. Similar to the intensity and latent heat trends, lower Sr compositions exhibit abrupt changes in tetragonality, whereas compositions above Sr$_{0.15}$ display increasingly gradual transitions with increasing Sr content. When the middle eigenvalue $\lambda_2$ of $\mathbf U$ approaches unity, the transformation becomes geometrically compatible, enabling coherent interface formation with reduced thermal hysteresis \cite{song2013}. Figure~\ref{fig:laue diffraction}(c) presents compatibility metric $\lvert\lambda_2 - 1\rvert$ and thermal hysteresis as functions of Sr composition. 

The composition Sr$_{0.19}$ exhibits the optimal lattice compatibility required for the ferroelectric to paraelectric phase transformation. At this composition, the thermal hysteresis reaches its minimum as $\lambda_{2} \to 1$. The latent heat analysis in Figure~\ref{fig:latent heat}(d) shows that Sr$_{0.19}$ falls within the transition region between first-order and second-order transformation behavior. Being located in this transitional regime, Sr$_{0.19}$ achieves nearly compatible phase transformations with minimal hysteresis and high reversibility. These characteristics make Sr$_{0.19}$ particularly suitable for pyroelectric energy conversion, as they jointly maximize the figure of merit, lattice compatibility, and ensure stable cyclic operation.

\noindent\textbf{Ferroelectric properties and energy conversion performance}

\begin{figure}[h]
\centering
\includegraphics[width=0.95\textwidth]{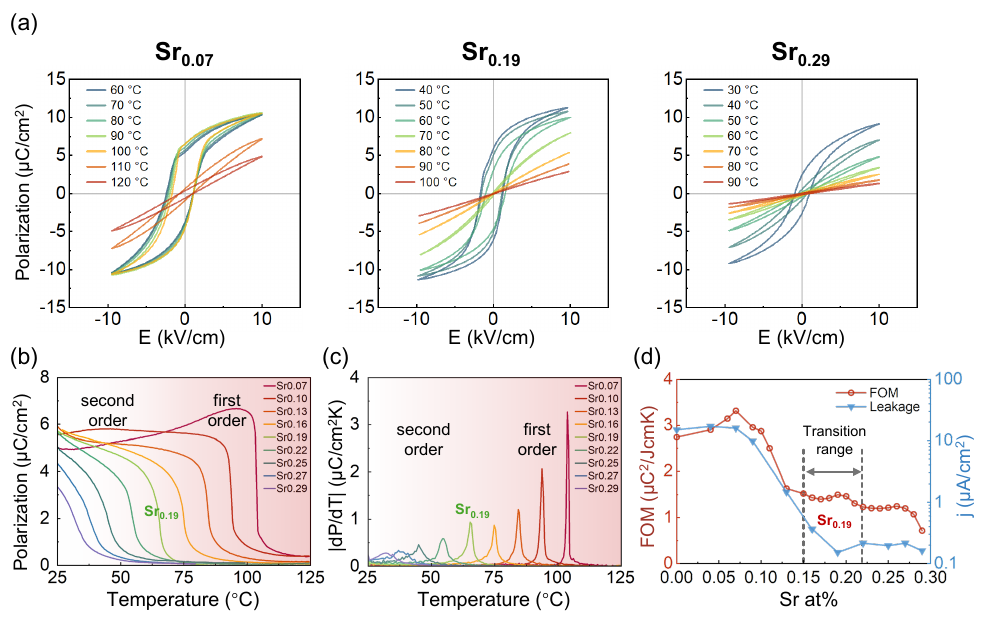}
\caption{\label{fig:ferro properties} (a) P-E characteristics for typical first-order Sr$_{0.07}$, transitioning Sr$_{0.19}$ and second-order Sr$_{0.29}$. (b) P-T and (c) $\lvert$dP/dT$\rvert$ values as a function of temperature for Sr$_{x}$, $0.07 \leq x \leq 0.29$. (d) Variations of energy conversion FOM and leakage current density with varying Sr compositions, characterized at their corresponding $T_c$.}
\end{figure}

The polarization–electric field (P–E) curves in Figure~\ref{fig:ferro properties}(a) reveal a systematic compositional evolution as the Sr content increases from 0 to 0.3. At low Sr compositions, for example Sr$_{0.07}$, the P–E loop exhibits large saturation polarization and remanent polarization with pronounced hysteresis, suggesting strong ferroelectric characteristics at room temperature. With increasing Sr content, both the saturation polarization and the remanent polarization decrease progressively, accompanied by a continuous narrowing of the hysteresis loop. At Sr$_{0.29}$, the loop becomes slim with substantially reduced remanent polarization.
The transport behavior during phase transformation is further illustrated by a series of temperature-dependent P–E curves, distinguished by different colors in Figure~\ref{fig:ferro properties}(a). As temperature increases, all three compositions undergo a ferroelectric to paraelectric phase transformation. For Sr$_{0.07}$ and Sr$_{0.29}$, the P–E loops remain open even at elevated temperatures, indicating charge dissipation in high temperature phase. In contrast, Sr$_{0.19}$ exhibits an ideal linear P–E response in the paraelectric phase with near-zero hysteresis, suggesting an ideal dielectric behavior with negligible dissipation during the transformation to the paraelectric state upon heating.

Figure~\ref{fig:ferro properties}(b) shows sharp polarization changes at transformation temperatures for lower Sr compositions, e.g. Sr$_{0.07}$ to Sr$_{0.13}$, characteristic of discontinuous first-order behavior. As Sr content increases beyond Sr$_{0.19}$, polarization changes become more gradual. Figure ~\ref{fig:ferro properties}(c) shows the pyroelectric coefficients, peaked at transformation temperatures. As expected, the compositions corresponding to first-order transformations exhibit large pyroelectric coefficient, which decreases as increasing Sr compositions. 

To evaluate the energy conversion performance by phase transformation, we calculated the figure of merit \cite{zhang2019power}:
\begin{equation}
    \text{FOM} = \frac{\Delta P \kappa}{\ell}
\end{equation}
where $\Delta P$ represents the polarization change during transformation, $\kappa = \left\vert \frac{\text{d}P}{\text{d}T} \right\vert_{\text{max}}$ denotes the maximum pyroelectric coefficient within the phase transition region, and $\ell$ is the latent heat of the phase transition.
Figure~\ref{fig:ferro properties}(d) presents the compositional dependence of FOM and electric leakage current density. For compositions below Sr$_{0.15}$ that exhibit first-order character, a maximum FOM of approximately 3.2 $\mu \text{C}^2$/(J cm K) occurs at Sr$_{0.07}$. However, this composition suffers from severe electric leakage, characterized by a leakage current density of 16.0 $\mu$A/cm$^2$ within the energy conversion temperature range, listed in Table~\ref{tab:leakage}. From structural analysis, it also shows a poor lattice compatibility with limited reversibility of energy conversion cycles.

\begin{table}[htb]
\caption{Representative energy conversion performance of the tested samples depending on phase transformation. $T_c$: Curie temperature; $j$: leakage current density; $\eta$: energy conversion efficiency. }
\label{tab:leakage}
\centering
\setlength{\tabcolsep}{2pt}
\renewcommand{\arraystretch}{1.15}
\begin{tabular}{c|c|c|c|c}
\toprule
Sample & $T_c~(^\circ\text{C})$ & FOM ($\mu$C$^2$/(J\ cm\ K)) &  $j$ $(\mathrm{\mu\text{A}/\text{cm}^2})$ & $\eta$ (\%) \\
\hline
Sr$_{0.07}$ & 102 & 3.31 & 16.02 & 1.8 \%\\
Sr$_{0.13}$ & 83 & 1.63 & 1.24  & 2.7 \% \\
\rowcolor{red!10}
Sr$_{0.19}$ & 64 & 1.50 & 0.15  & 5.5 \%\\
Sr$_{0.25}$ & 45 & 1.22 & 0.20  & 5.6 \% \\
Sr$_{0.29}$ & 35 & 0.72 & 0.16   & 1.3 \%\\
\bottomrule
\end{tabular}
\end{table}

The sample Sr$_{0.19}$ achieves a high figure of merit (1.5 $\mu \text{C}^2$/(J cm K)), almost equivalent to the single crystal performance among reported phase-transforming ferroelectric materials \cite{Moya2013, bucsek2020energy}. In this very low–temperature regime, unlike thermoelectric or Rankine-cycle systems whose performance drops sharply below 100$~^\circ$C (especially below 70$~^\circ$C), Sr$_{0.19}$ maintains promising energy harvesting from waste heat around 64$~^\circ$C.
More importantly, Sr$_{0.19}$ maintains a suppressed leakage current density of only 0.15 $\mu$A/cm$^2$, more than two orders of magnitude lower than Sr$_{0.07}$, given in Table \ref{tab:leakage}. Relative to second-order featured compositions, Sr$_{0.19}$ retains a high performance in terms of FOM, exceeding those of Sr$_{0.25}$ and Sr$_{0.29}$.

The thermodynamic efficiency, defined as the ratio of generated electrostatic energy density to the absorbed heat during phase transformation, is listed in Table~\ref{tab:leakage}. The Sr$_{0.19}$ sample exhibited a promising thermodynamic efficiency of 5.5\% while closely satisfying lattice compatibility, thereby enhancing phase reversibility upon thermal cycling. The synergistic optimization of low leakage, high energy density, and high efficiency makes Sr$_{0.19}$ the most suitable candidate for practical energy harvesting applications. Detailed calculations of electrostatic energy density and efficiency for the ferroelectric-to-paraelectric phase transformation are provided in the Supplemental Materials.

\noindent\textbf{Energy conversion by first-order, transitional and second-order phase transforming Ba$_{1-x}$Sr$_x$TiO$_3$}

\begin{figure}[htb]
\centering
\includegraphics[width=0.75\textwidth]{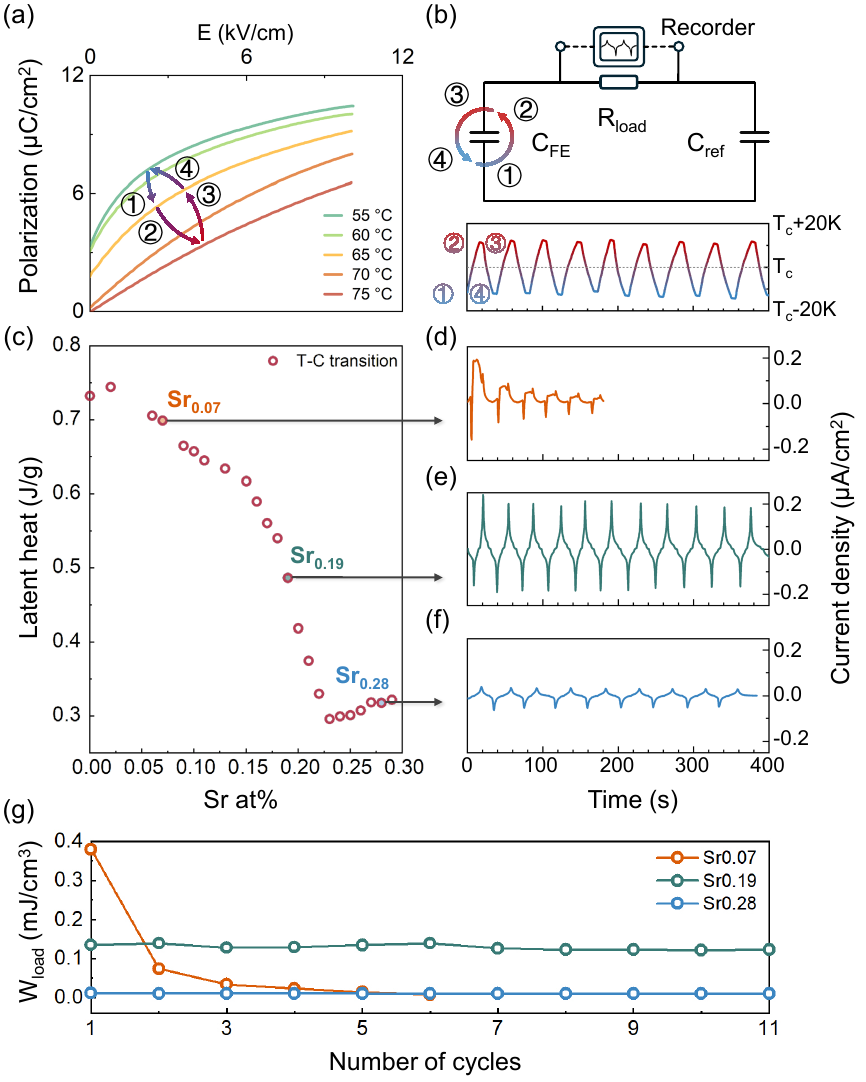}
\caption{\label{fig:demonstration} (a) Polarization response of Sr$_{0.19}$ sample under positive electric fields up to 10 kV/cm at temperatures ranging from 65$~^\circ$C to 75$~^\circ$C. (b) Schematic of the energy conversion demonstration circuit with temperature controlled varying around $T_c$. \textcircled{1}-\textcircled{4} represents four stages of the thermodynamic cycle: \textcircled{1}: heating from low temperature to $T_c$; \textcircled{2}: heating from $T_c$ to high temperature; \textcircled{3}: cooling from high temperature to $T_c$; \textcircled{4}: cooling from $T_c$ to low temperature. (c) Latent heat of tetragonal-cubic phase transformation in BST material system and energy conversion demonstration results of (d) Sr$_{0.07}$ (e) Sr$_{0.19}$ and (f) Sr$_{0.28}$ bulk ceramic samples. (g) Work done on the 1~M$\Omega$ load resistor over 11 cycles of Sr$_{0.07}$, Sr$_{0.19}$ and Sr$_{0.28}$ bulk ceramic samples. }
\end{figure}
 
Energy-conversion demonstrations were carried out  through the thermodynamic cycle 1–2–3–4, as illustrated in Figure~\ref{fig:demonstration}(a). Figure~\ref{fig:demonstration}(b) shows the electric circuit used to collect the electric energy generated by temperature fluctuations of approximately $\pm 15$~K around the transformation temperature $T_c$, which drive phase transformations in the samples under a fully power-source-free condition. The corresponding circuit configuration followed Ref.~\cite{zhang2019power}. In this setup, C$_\text{FE}$ denotes the phase-transforming capacitor employing Sr$_x$ as the dielectric layer, while a non-transforming capacitor C$_\text{ref}$ serves as a charge reservoir during the charging and recharging processes. The load resistor was selected to be 1~M$\Omega$.

Corresponding to the Sr compositions selected from the latent heat diagram in Figure~\ref{fig:demonstration}(c), Figure~\ref{fig:demonstration}(d)–(f) shows the generated current density over multiple thermal cycles. The Sr$_{0.07}$ capacitor, representative of a first-order transformation, produces an initial maximum current density of 0.2~$\mu$A/cm$^2$ but degrades rapidly by 80\% due to increased leakage associated with poor lattice compatibility. The Sr$_{0.29}$ capacitor, exhibiting a second-order transformation, generates a much lower current density of approximately 0.03~$\mu$A/cm$^2$ that remains relatively stable with minimal degradation. In contrast, the Sr$_{0.19}$ capacitor with transitional transformation character achieves a current density of approximately 0.2~$\mu$A/cm$^2$ at 64$~^\circ$C while maintaining stable phase reversibility throughout cycling. This prolonged stability combined with a practical level of current generation distinguishes Sr$_{0.19}$ from both typical first-order and second-order compositions and motivates its use in further device optimization.

\noindent\textbf{Energy conversion demonstration by phase transforming ferroelectric MLCC}

\begin{figure}
    \centering
    \includegraphics[width=0.9\textwidth]{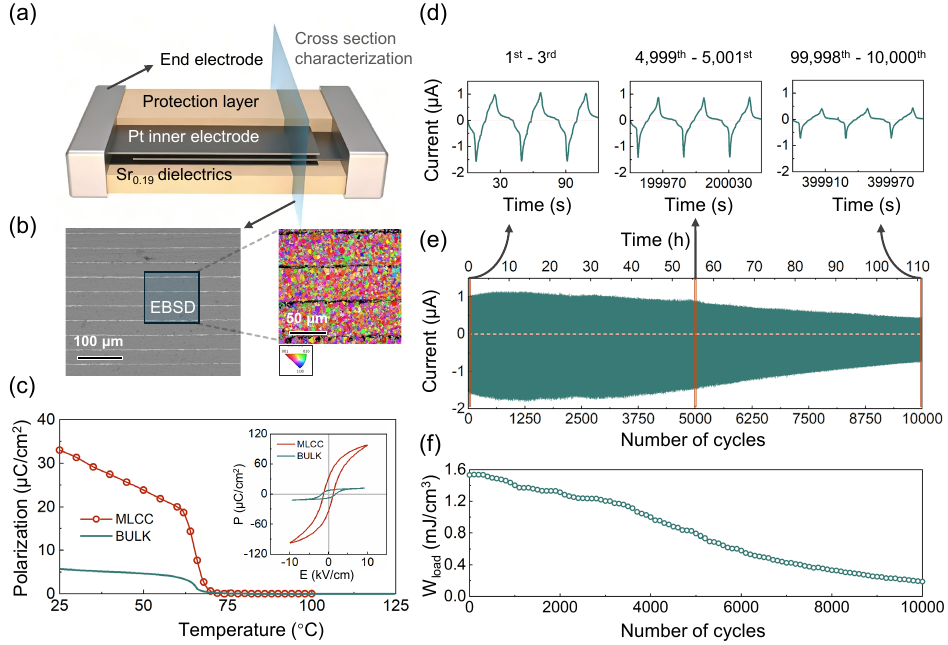}
    \caption{(a) Schematic of the Sr$_{0.19}$ MLCC device structure. (b) Cross sectional SEM and EBSD images of the Sr$_{0.19}$ MLCC device. (c) P-T curves of Sr$_{0.19}$ MLCC and bulk sample with room-temperature P-E loops . (d) The current output details and (f) overview of Sr$_{0.19}$ MLCC during long-cycle energy conversion demonstration. (g) Work done on the 1~M$\Omega$ load resistor over 10,000 cycles of Sr$_{0.19}$ MLCC. }
    \label{fig:mlcc}
\end{figure}

The interval between Sr$_{0.15}$ and Sr$_{0.22}$ corresponds to a transition from first-order to second-order transformation character, providing a balance between fatigue resistance and promising energy conversion performance. Based on these observations, we designed a multilayer planar capacitor (MLCC) using Sr$_{0.19}$ as the dielectric layer, with an active area of 20.68~mm$^{2}$, a total thickness of 0.5~mm, and 9 dielectric layers of effective thickness $\sim$50~$\mu$m (see Methods for fabrication details). Figure~\ref{fig:mlcc}(a) and (b) illustrate the design of the MLCC device, fabricated from multilayered Sr$_{0.19}$ dielectric materials. As shown, the microstructure of the dielectric layers is dense without nominal pores, and the grain morphology is homogeneous with a randomized distribution of orientations. The average grain size is approximately 5~$\mu$m, maintaining sufficient coherent interfaces between adjacent layers. 

Figure~\ref{fig:mlcc}(c) characterizes the P–E and P–T behaviors of the 9 layer Sr$_{0.19}$ MLCC. Compared with its bulk counterpart, the MLCC device exhibits a substantial enhancement in both saturation polarization and remanent polarization. Consequently, the polarization jump during phase transformation increases by approximately a factor of five, directly amplifying the extractable electrical work per thermodynamic cycle.

The energy conversion performance is presented in Figure~\ref{fig:mlcc}(d) and (e). The Sr$_{0.19}$ MLCC device sustains a stable pyroelectric current of $0.7\sim 1.6~\mu$A over 10,000 consecutive thermal cycles, corresponding to about 110~h of continuous operation, which is approximately one week. Notably, the device operates at a low grade heat regime of only 64$~^\circ$C, demonstrating continuous and stable heat to electricity conversion under mild thermal conditions that are readily accessible in practical waste heat environments. The current traces recorded at the initial stage, after 5,000 cycles, and after 10,000 cycles exhibit minimal variation, confirming excellent operational stability. Figure~\ref{fig:mlcc}(f) shows the work delivered to a 1~M$\Omega$ load resistor by the Sr$_{0.19}$ MLCC energy converter over 10,000 thermodynamic cycles. During the first several hundred cycles, the average work output is approximately 1.6~mJ/cm$^3$. With continued cycling, the output gradually decreases to one half of this value by 5,000 cycles and to approximately one quarter by 10,000 cycles, while maintaining sustained energy generation throughout the week long operation at 64$~^\circ$C.

\begin{figure}
    \centering
    \includegraphics[width=0.45\textwidth]{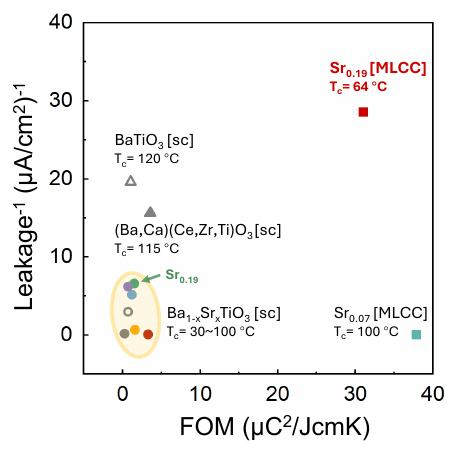}
    \caption{Leakage versus figure of merit in Sr doped barium titanate systems and similar systems \cite{Moya2013, zhang2021energy}. The Sr$_{0.19}$ MLCC stands out by simultaneously achieving high FOM and low leakage.}
    \label{fig:benchmark}
\end{figure}

Figure~\ref{fig:benchmark} maps leakage current density ($\mu$A/cm$^{2}$) against figure of merit (FOM) for representative BaTiO$_3$-based systems in both bulk and MLCC configurations. The bulk single crystal BaTiO$_3$ exhibits lower leakage than its polycrystalline counterparts, yet remains confined to a relatively modest FOM regime. Across the bulk compositions, including (Ba,Ca)(Ce,Zr,Ti)O$_3$ and Ba$_{1-x}$Sr$_x$TiO$_3$, a clear trade-off is observed: enhancement of polarization-related FOM generally accompanies increased leakage, which ultimately constrains effective energy conversion.

In sharp contrast, the Sr$_{0.19}$ MLCC distinctly deviates from this trend. It occupies the upper-right performance window, simultaneously achieving a substantially higher FOM while maintaining low leakage compared with other high-FOM compositions. This separation from the conventional trade-off line indicates that the multilayer capacitor architecture fundamentally alters the performance landscape. Rather than merely improving polarization response, the MLCC configuration amplifies the phase-transformation-induced polarization jump while concurrently suppressing electrical dissipation. As a result, Sr$_{0.19}$ MLCC establishes a superior balance between energy output and operational stability, marking a clear advancement over bulk counterparts and other substituted BaTiO$_3$ systems.

\noindent\textbf{Conclusion}

We uncovered a compositionally-driven transition from first-order to second-order phase transformation behavior in the Ba$_{1-x}$Sr$_x$TiO$_3$ system. Through comprehensive thermal analysis, synchrotron X-ray diffraction, and ferroelectric characterization, we identify a transitional regime between Sr$_{0.15}$ and Sr$_{0.22}$ where the transformation mechanism fundamentally evolves. The latent heat decreases by a factor of two across this compositional window, although phase boundaries remain continuous and linear in the temperature-composition phase diagram.

The composition Sr$_{0.19}$ within this transition region closely satisfies lattice compatibility for the cubic-to-tetragonal transformation, as evidenced by minimum in both $|\lambda_2 - 1|$ and thermal hysteresis. The optimal compatibility enables enhanced phase reversibility and suppresses electrical leakage by more than two orders of magnitude compared to strong first-order compositions. Energy conversion demonstrations confirm that one Sr$_{0.19}$ MLCC converter delivers pyroelectric current of 1.6 $\mu$A with stable cycling performance at 64$~^\circ$C over long operation without external bias field.

These results demonstrate that targeting transitional compositions between first-order and second-order behavior provides a viable strategy for optimizing pyroelectric energy conversion materials. Rather than maximizing individual properties such as pyroelectric coefficient or minimizing hysteresis independently, compositions in the transition region balance substantial polarization response necessary for energy harvesting with operational reliability required for practical devices. This materials design strategy extends pyroelectric energy harvesting to low-grade waste heat sources through compositional tuning of phase transformation character.

\begin{acknowledgments}
R. G., X. H. and X.C. thank the financial support under GRF Grants No. 16203021, 16204022 and No. 16203023 by Research Grants Council, Hong Kong. This research used resources of the Advanced Light Source, a U.S. DOE Office of Science User Facility under contract no. DE-AC02-05CH11231. C. Z. acknowledge the support by National Natural Science Foundation of China (No.12204350)
\end{acknowledgments}

\bibliographystyle{unsrt}
\bibliography{cas-refs}

\noindent\textbf{Methods.}

\noindent\textbf{Materials synthesis.}
Strontium-doped barium titanate (BST) ceramics were synthesized through a conventional solid-state reaction method. High-purity BaCO$_3$ (99.99\%, Alfa Aesar), TiO$_2$ (99.9\%, Sigma-Aldrich), and SrCO$_3$ (99.99\%, Aladdin) powders were mixed according to the stoichiometric formula Ba$_{1-x}$Sr$_x$TiO$_3$ with $x$ ranging from 0 to 0.3 in increments of 0.01. Samples are designated by their nominal Sr content as Sr$_x$.

The mixed powders were dispersed in ethanol and homogenized by ball milling with zirconia media using a MITR YXQM-1L planetary ball mill for 20 hours at 300 rpm. The milled powders were subsequently calcined at 1000$~^\circ$C for 6 hours in a muffle furnace (TMAXCN) to promote solid-state reaction and compound formation. Following calcination, the powders underwent secondary ball milling and granulation to improve flowability and achieve uniform particle size distribution.

The calcined powders were pressed using a hydraulic press (MITR MC-15) to form green bodies with dimensions of 10 mm diameter and 1.5 mm thickness. Sintering was performed in a muffle furnace (TMAXCN) under ambient atmosphere at 1350$~^\circ$C for 9 hours to obtain dense BST ceramic samples.

The Sr$_{0.19}$ multilayer ceramic capacitor (MLCC) was fabricated using a standard tape-casting process. Calcined Sr$_{0.19}$ powder was dispersed into a slurry, cast into green tapes, and dried to obtain uniform dielectric films. The films were punched to the target area, printed with internal electrodes, and stacked to form a 9-layer architecture with an effective dielectric thickness of approximately 50~$\mu$m per layer. The laminate was pressed, subjected to binder burnout, and co-sintered under controlled conditions to produce a dense multilayer structure suitable for pyroelectric energy-conversion testing.

\noindent\textbf{Electron backscatter diffraction.}
The sintered specimens were mechanically polished and characterized by electron backscatter diffraction (EBSD) using a JEOL 7800 scanning electron microscope to reveal the microstructure and grain morphology for the room temperature phases. The orientation map was obtained by indexing each of the Kikuchi patterns by tetragonal symmetry.

\noindent\textbf{Differential scanning calorimetry.}
Differential scanning calorimetry (DSC, TA Instruments DSC250) measurements were performed on all synthesized samples to investigate phase transformation behavior. Measurements were conducted over the temperature range of $-70~^\circ$C to $140~^\circ$C to capture both cubic-to-tetragonal and tetragonal-to-orthorhombic transitions. The latent heat for each phase transformation was calculated from the integrated area of the heat flow peaks.

\noindent\textbf{Synchrotron X-ray diffraction.}
Synchrotron X-ray diffraction (SR-XRD) measurements were performed at Beamline 12.3.2, Advanced Light Source, Lawrence Berkeley National Laboratory. For the cubic-to-tetragonal transformation, Laue diffraction patterns were captured at temperatures before, during, and after the transformation. Temperature-dependent intensity of characteristic diffraction peaks was tracked and normalized with respect to the cubic phase. Monochromatic energy scans were performed to determine the lattice parameters of the cubic (space group $Pm\bar{3}m$), tetragonal (space group $P4mm$), and orthorhombic (space group $Amm2$) phases for each composition. The calculated lattice parameters are provided in the Supplemental Materials.

\noindent\textbf{Ferroelectric characterization.}
Ferroelectric properties were characterized using an AixACCT TF 2000E analyzer. Specimens were mechanically polished with 400 mesh and 800 mesh sandpaper, then coated with conductive silver paste electrodes with an effective surface area of approximately 55 mm$^2$ and thickness of 0.6 mm. Measurements were conducted under an applied electric field of 10 Hz and 10 kV/cm, with remnant polarization recorded as a function of temperature. The pyroelectric coefficient $\lvert\text{d}P/\text{d}T\rvert$ was calculated through numerical differentiation of the polarization-temperature profile.

\end{document}